\documentclass[twocolumn]{aastex631}

\newcommand\qpah{q_\mathrm{PAH}}

\usepackage{amsmath}

\begin{document}

\title[PAHs and R(V) variation]{Dust extinction-curve variation in the translucent interstellar medium is driven by PAH growth}

\author[0000-0003-3112-3305]{Xiangyu Zhang}
\affiliation{Max Planck Institute for Astronomy, Königstuhl 17, 69117 Heidelberg, Germany}

\author[0000-0001-7449-4638]{Brandon S. Hensley}
\affiliation{Jet Propulsion Laboratory, California Institute of Technology, 4800 Oak Grove Drive, Pasadena, CA 91109, USA}

\author[0000-0001-5417-2260]{Gregory M. Green}
\affiliation{Max Planck Institute for Astronomy, Königstuhl 17, 69117 Heidelberg, Germany}

\begin{abstract}
The first all-sky, high-resolution, 3D map of the optical extinction curve of the Milky Way \citep{ZG24a} revealed an unexpected steepening of the extinction curve in the moderate-density, ``translucent'' interstellar medium (ISM). We argue that this trend is driven by growth of polycyclic aromatic hydrocarbons (PAHs) through gas-phase accretion. We find a strong anti-correlation between the slope of the optical extinction curve -- parameterized by $R(V)$ -- and maps of PAH abundance -- parameterized by $q_{\rm PAH}$ -- derived from infrared emission. The range of observed $q_{\rm PAH}$ indicates PAH growth by a factor of $\sim$2 between $A_V \simeq 1$ and 3. This implies a factor-of-two stronger 2175\,\AA\ feature, which is sufficient to lower $R(V)$ by the observed amount. This level of PAH growth is possible given rapid accretion timescales and the depletion of carbon in the translucent ISM. Spectral observations by JWST would provide a definitive test of this proposed explanation of $R(V)$ variation.
\end{abstract}


\section{Introduction}

Extinction from dust as a function of wavelength, the ``extinction curve,'' is a key observable to constrain the composition and evolution of dust. At optical through near-infrared (NIR) wavelengths, extinction-curve variation is usually described by a single parameter, $R(V)\equiv A(V)/E(B-V)$ \citep{CCM89}. Larger values of $R(V)$ indicate a flatter extinction curve, which is generally thought to indicate a greater proportion of larger dust grains \citep[e.g.,][]{Weingartner:2001}. Lower values indicate that the extinction curve falls off more steeply with increasing wavelength, which generally indicates a larger proportion of small dust grains. While smaller variations in the optical-NIR extinction curve of Milky Way dust exist (e.g., \citealt{Massa20_bumps},
\citealt{Maiz_7700}, \citealt[no relation]{Ruoyi24_ISS}), the dominant mode of variation is captured by $R(V)$. Parametrization by $R(V)$ has proven an effective model of the general profile of optical-NIR extinction curves in the Milky Way (e.g.,~\citealt{Gordon23_Rv, Fitzpatrick19_Rv_deviation, Schlafly16_ExtCurve}).

Despite the success of the phenomenological parametrization by $R(V)$, the detailed physical causes of $R(V)$ variation have remained unclear. The largest challenge has been the limited number of lines of sight along which $R(V)$ has been determined. This limited data volume makes it challenging to correlate $R(V)$ with other physical parameters. Measurements of $R(V)$ have historically been limited to a few hundred lines of sight towards extincted O- and B-stars \citep[e.g., ][]{Valencic04_ext_curves, Fitzpatrick07_ext_curves} . 

New, large spectroscopic and multiband photometric surveys have made it possible to expand the number of $R(V)$ measurements by orders of magnitude. \citet{Schlafly16_ExtCurve} 
determined $R(V)$ toward $\sim37,000$ stars and found a strong anti-correlation between $R(V)$ and the spectral index of the far-infrared dust emissivity power law, $\beta$. \citet[no relation]{ZYC23_Rv_map} reliably determined dust $R(V)$ to over 1 million stars and found an anti-correlation between $R(V)$ and $\beta$, $N_\mathrm{H_2}$, $N_\mathrm{H_2}/N_\mathrm{HI}$ and the gas-to-dust ratio, and a positive correlation between $R(V)$ and $N_{\rm HI}$. 

\citet[``ZG24'']{ZG24a} used \textit{Gaia} XP spectra \citep{GaiaMission,GaiaDR3,GaiaXPProcessingValidation,GaiaXPExternalValidation} to produce the largest catalog (over 130 million stars) of $R(V)$ in the Milky Way and Magellanic clouds at the time of writing. Contrary to previous expectations that grain growth should cause $R(V)$ to increase as ISM density increases, ZG24 find that $R(V)$ tends to decrease with $A(V)$ in the translucent ISM, before increasing again in the dense molecular ISM. Here, we seek to explain this observed decrease in $R(V)$ in the translucent ISM.

Polycyclic aromatic hydrocarbons (PAHs) are one of the most common classes of organic molecules in the universe. Their characteristic emission features at 3.3, 6.2, 7.7, 8.6, 11.2, 12.7, and 16.4\,$\mathrm{\mu m}$ have been observed in various sources in the Milky Way and other galaxies \citep{Tielens08_PAH_review}. Although questions remain in the formation of PAHs \citep{LiAigen20_PAH_review}, theories, observations and experiments suggest that growth of PAH in the ISM is viable (e.g. \citealt{Krasnokutski_PAH_accretion, McGuire21_PAH_detection, Jones11_PAH_formation, Parler12_naphthalene_formation}). 

PAHs have been proposed as the carrier of the 2175\,\AA{} bump, the most prominent feature in the UV region of the extinction curve \citep{Joblin92_PAH, Li_Draine01_PAH, Malloci04_PAH, Malloci08_PAH, Cecchi-Pestellini_PAH, Steglich10_PAH, Steglich12_PAH}. An anti-correlation between the strength of the 2175\,\AA{} feature and $R(V)$ has long been recognized (e.g.~\citealt{CCM89}), but its physical explanation has been unclear.

In this work, we propose that $R(V)$ variation in translucent clouds is dominated by the variation of PAH abundance. As PAH mass increases, the strength of 2175\,\AA{} feature is enhanced, and the red wing of the feature is steep enough in optical-NIR to change $R(V)$, which explains the following phenomena:
\begin{itemize}
\vspace{-0.3cm}
  \item Anti-correlation between $R(V)$ and the 2175\,\AA{} feature;

\vspace{-0.3cm}
  \item Decreasing $R(V)$ as $A(V)$ increases in the translucent ISM.
  \vspace{-0.3cm}
\end{itemize}
In addition, if $R(V)$ is driven by PAH abundance, variation of $R(V)$ will serve as an indicator of the growth of PAHs in the ISM, which provides important implications for future James Webb Space Telescope (JWST) observations.

This paper is organized as follows. In Section~\ref{sec:obs}, we describe the observed anti-correlation between $R(V)$ and PAH abundance by comparing 2D $R(V)$ maps and a PAH abundance ratio ($\qpah$) map. In Section~\ref{sec:theory}, we demonstrate the theoretical viability of PAH growth as a driver of $R(V)$ variation, and rule out the possibility of accretion onto larger dust grains having a significant effect on $R(V)$ in this regime. In Section~\ref{sec:discussion}, we discuss the implications of our hypothesis for PAH growth and future observations with JWST.

\section{An Anti-correlation between $R(V)$ and PAH mass ratio}\label{sec:obs}
In the regions where $R(V)$ decreases with $A(V)$ in ZG24, we generally find an anti-correlation between $q_\mathrm{PAH}$ and $R(V)$ from the maps of \citet{qPAHmap_planck} and \cite{ZG24a}. This relation persists among the translucent regions in the vicinity of various molecular clouds (Aquila S, California, Orion A, Perseus, Ursa MA, etc) in the Milky Way and the Large Magellanic Cloud (LMC).

An ideal comparison between $R(V)$ and PAH abundance would be conducted in 3D. However, PAH maps are based on emission features of PAHs integrated to infinite distance, and are thus 2D in nature. To measure the same column of dust, we build 2D $R(V)$ maps by a binning method, using only stars that are behind most of the dust cloud. We describe the details of the method, as well as the selection of stars, in Appendix~\ref{app:binning}.

We use the PAH mass ratio ($q_\mathrm{PAH}$) map from Planck intermediate results \citep{qPAHmap_planck}, which models dust emission data from Planck \citep{Planck_01, Planck_08}, IRAS \citep{IRIS}, and WISE \citep{WISE} infrared observations with the dust model of \citet{DL07}. The fitting of 2D emission data assumes that the radiation field and the temperature of the dust along the line of sight are the same, which does not necessarily hold for lines of sight that cross multiple dust clouds or extended dust structures. Therefore, we select lines of sight where the majority of the extinction (over 70\%) is concentrated within a 500-pc distance range: 

\begin{equation}\label{eqn:multi_cloud}
\max_{0<d_0<4.5 \mathrm{kpc}} \{\int_{d_0}^{d_0+500\,\mathrm{pc}}dE \}/E_\mathrm{5\,kpc}>0.7 .
\end{equation}
\newline

We present $A(V)$, $R(V)$ and $q_\mathrm{PAH}$ of five clouds in the Milky Way in Figures~\ref{fig:Av_Rv_qPAH} and \ref{fig:Av_Rv_qPAH_continuation}, in which most sightlines fulfill our requirements. Despite finer structures in both $R(V)$ and $q_\mathrm{PAH}$ maps, we generally observe lower $R(V)$ in higher-$q_\mathrm{PAH}$ regions in all five clouds, which is shown quantitatively in Figure~\ref{fig:quant_Rv_qPAH}. 
An increase in $\qpah$ from 4\% to 8\% is observed to result in a decrease in $R(V)$ by $\sim 0.2-0.5$.
The median $R(V)$ at low $q_\mathrm{PAH}$ varies among the clouds, which may be attributed to their different initial dust grain size distribution, as well as magnetic field variations \citep{Voshchinnikov:1989}. In all five clouds, $R(V)$ approaches a value of $~\sim2.9$ at $q_\mathrm{PAH}\simeq8\%$. 

The scatter of $R(V)$ at given $q_\mathrm{PAH}$ is large ($\sim0.3\,\mathrm{mag}$) in some clouds shown in Figures~\ref{fig:Av_Rv_qPAH} and \ref{fig:Av_Rv_qPAH_continuation}. The following factors may contribute to the scatter:
\begin{itemize}
\vspace{-0.3cm}
\item Finer structures of clouds exist but the $q_\mathrm{PAH}$ map has limited angular resolution.
\vspace{-0.3cm}
\item Variation of both $q_\mathrm{PAH}$ and $R(V)$ exist along the line of sight, which 2D maps cannot represent.
\vspace{-0.3cm}
\item Uncertainty of measured $R(V)$ of individual stars.
\vspace{-0.3cm}
\item The $W_3$ photometric band has an underlying continuum that is not from PAHs, and the decomposition provided by the DL07 model may not be perfect. Further, a number of gas emission lines can contribute to the total $W_3$ flux.
\vspace{-0.3cm}
\item Magnetic fields may vary in the dust clouds, which can lead to $R(V)$ variation of order $\sim 0.2$ \citep{Voshchinnikov:1989}.
\vspace{-0.3cm}
\end{itemize}

\begin{figure*}
    \centering
    \includegraphics[width=1.0\linewidth]{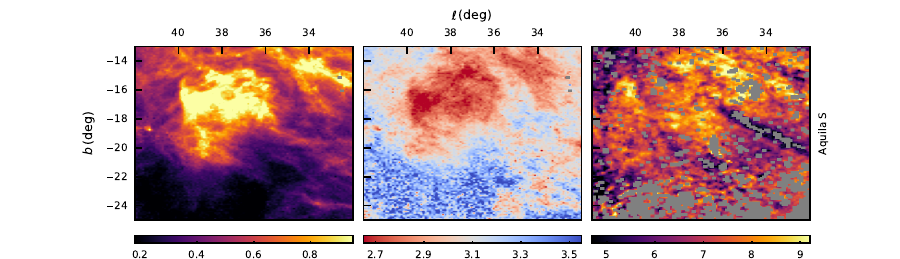}
    \includegraphics[width=1.0\linewidth]{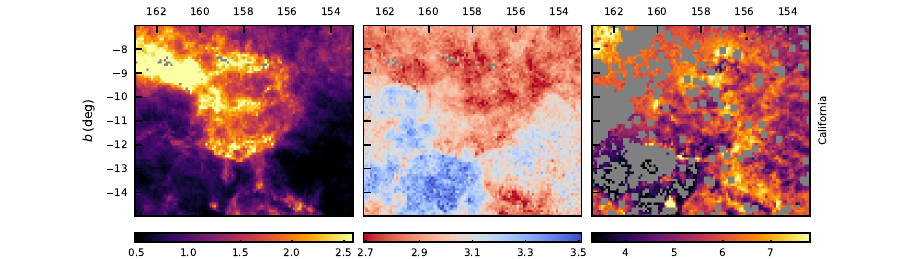}
    \includegraphics[width=1.0\linewidth]{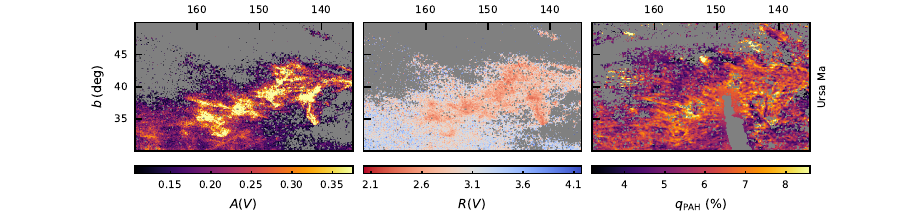}
    \caption{Comparison of 2D distribution of $A(V)$, $R(V)$, $q_\mathrm{PAH}$ for dust clouds. For $A(V)$ and $R(V)$ maps, we select high-quality stars from \cite{ZG24a}, and remove lines of sight where only partial extinction is covered by XP. For $q_\mathrm{PAH}$ map, we remove lines of sight where the $q_\mathrm{PAH}$ map is of low quality ($\chi^2/\mathrm{DOF}>3$ or $q_\mathrm{PAH}/\sigma(q_\mathrm{PAH})<3$), or that cross multiple clouds (Equation~\ref{eqn:multi_cloud}). The binning method and selection of stars in the 2D maps are described in Appendix~\ref{app:binning}. Despite different baselines and finer structures in different maps, all clouds tend to have lower $R(V)$ at high-$q_\mathrm{PAH}$ regions, as indicated quantitatively on the per-star basis in Figure~\ref{fig:quant_Rv_qPAH}}
    \label{fig:Av_Rv_qPAH}
\end{figure*}

\begin{figure*}
    \centering
    \includegraphics[width=1.0\linewidth]{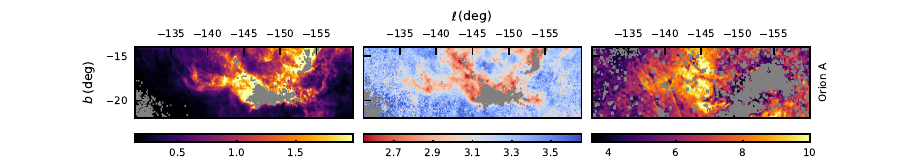}
    \includegraphics[width=1.0\linewidth]{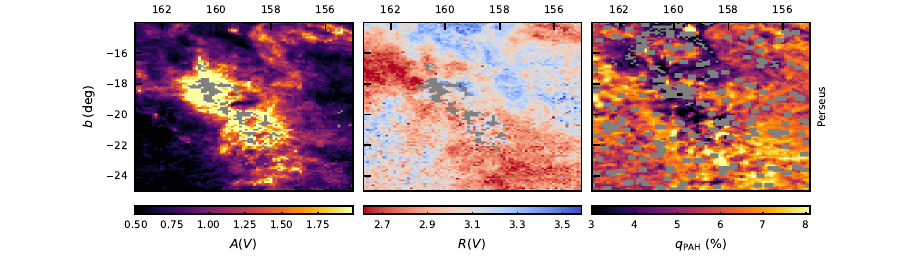}
    \caption{Continuation of Figure~\ref{fig:Av_Rv_qPAH}}
    \label{fig:Av_Rv_qPAH_continuation}
\end{figure*}

\begin{figure*}
    \includegraphics[width=1.0\linewidth]{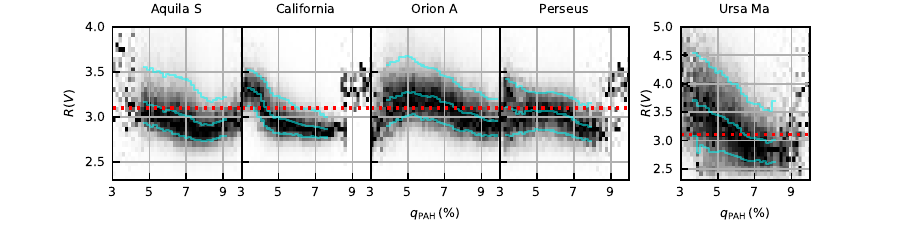}
    \caption{Correlation between $R(V)$ and $q_\mathrm{PAH}$ using stars that pass all the cuts in Figure~\ref{fig:Av_Rv_qPAH}. 16th, 50th and 84th percentiles are in cyan lines. We also label $R(V)=3.1$, the Milky Way average value, in red dotted lines.  Although different clouds have different baselines of $R(V)$, due to different initial distribution of dust grain size, all dust clouds show anti-correlation between $R(V)$ and $q_\mathrm{PAH}$. }
    \label{fig:quant_Rv_qPAH}
\end{figure*}

\begin{figure*}
    \centering
    \includegraphics[width=0.8\linewidth]{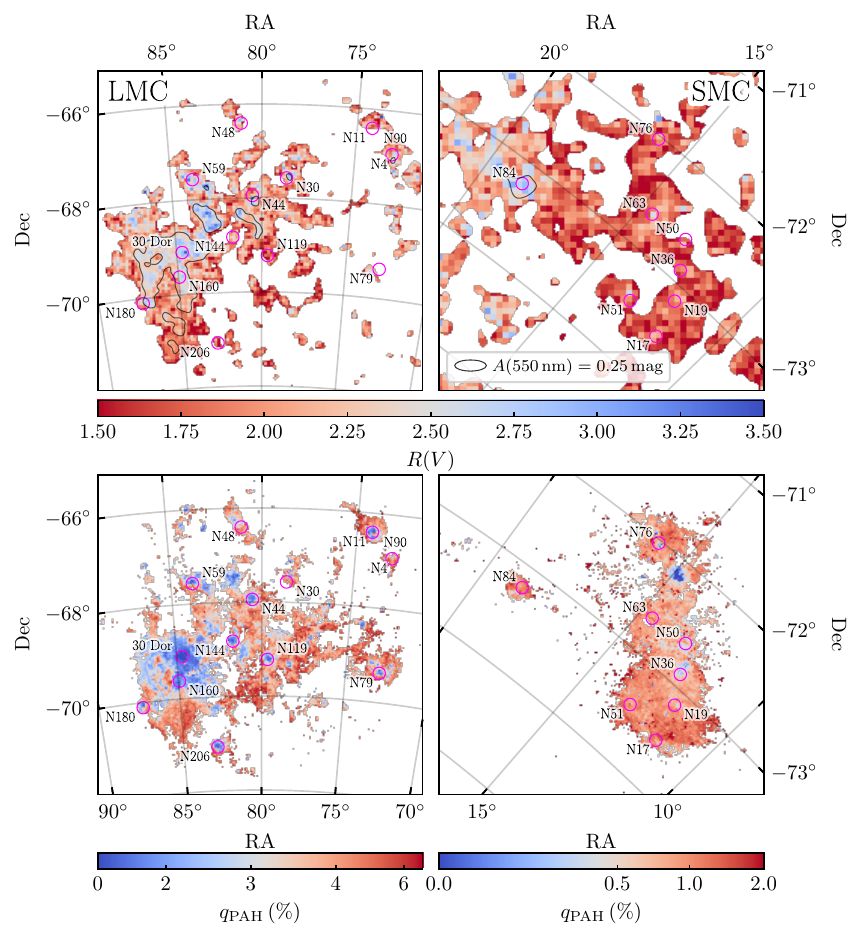}
    \caption{$R(V)$ distribution of LMC and SMC. We select stars by distance and proper motion and remove foreground extinction, described in appendix~\ref{app:MC_binning},  We label a few dense dust clouds indicated by H~{II} signal from \cite{Lopez14_HII}, which tend to higher $R(V)$ than their neighborhood. We also show the $q_\mathrm{PAH}$ map from \cite{Chastenet19}. The H~{II} regions also have lower $q_\mathrm{PAH}$. This anti-correlation can be explained as the destruction of PAHs by coagulation in dense dust clouds or UV photons in star-forming regions. }
    \label{fig:MC}
\end{figure*}

In addition to the Milky Way, the anti-correlation between $q_\mathrm{PAH}$ and $R(V)$ is also observed in the LMC. Figure~\ref{fig:MC} shows the 2D $R(V)$ distribution from ZG24 and the $q_\mathrm{PAH}$ map from \cite{Chastenet19}. Foreground dust of LMC and SMC is not negligible, so we exclude the contamination from the disk and the halo of the Milky Way. We describe the detailed method and foreground removal in Appendix~\ref{app:MC_binning}. In the LMC, most H\,{II} regions have lower $q_\mathrm{PAH}$ and higher $R(V)$ than their vicinity, which is consistent with the trend we find in the Milky Way. This can be attributed to the destruction of PAHs by UV photons or coagulation. We also find anti-correlation on a per-star basis in the LMC, as illustrated in Figure~\ref{fig:quant_Rv_qPAH}. 

In the SMC, $R(V)$ and $q_\mathrm{PAH}$ appear to be independent. It is notable that the PAH mass ratio, as well as its variation, is generally low in the SMC, which is not enough to change $R(V)$ substantially. The size distribution of larger grains could also be significantly different from that of the Milky Way \citep{Weingartner:2001}. 

\begin{figure}
    \centering
    \includegraphics[width=1\linewidth]{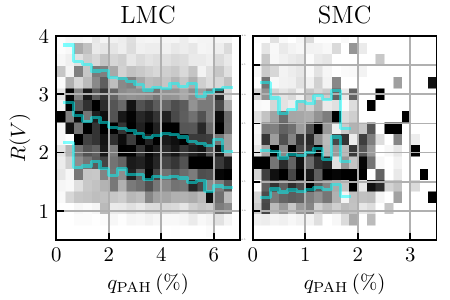}
    \caption{Quantitative comparison on the per-star basis between $R(V)$ and $q_\mathrm{PAH}$ in LMC (left panel) and SMC (right panel). The median value of $R(V)$ decreases by $0.4$ as $q_\mathrm{PAH}$ increases from $1\%$ to $6\%$ in the LMC. In the SMC, the trend is not clear because of small variation range of $q_\mathrm{PAH}$ as well as the difference in spatial coverage of both maps. 
    }
    \label{fig:quant_Rv_qPAH}
\end{figure}

\section{Theoretical explanation of $R(V)$ variation in the translucent ISM}\label{sec:theory}

In this section, we demonstrate theoretically that the variation of $\qpah$ is capable of changing $R(V)$ as we observe in section~\ref{sec:obs}, and rule out the possibility of accretion onto larger dust grains causing the observed decrease in $R(V)$.

We assume that dust grains are well mixed when a cloud initially forms, such that the initial grain size distribution (of each dust component) is homogeneous. As a cloud evolves, growth and destruction of dust grains shape the size distribution differently at different densities within the cloud, which gives rise to $A(V)$ and $R(V)$ variations.  

We adopt the dust grain model by \citet{Astrodust_PAH}, which consists of two components, ``astrodust'' \citep{Astrodust_dielectric}, and PAHs. Astrodust is a composite material to account for large grains ($\gtrsim 500 \AA$), which contains the majority of the elements depleted from the gas phase, such as C, O, Mg, Si and Fe. Small grains are accounted for by PAHs.
By fitting the grain-size distribution of astrodust and PAHs, this model reproduces the observed extinction, polarized extinction, emission, and polarized emission from dust in the diffuse ISM. We use the best-fit grain-size distribution of astrodust and PAHs in \citet{Astrodust_PAH} as the initial distribution, and then model the growth of dust through accretion.

In the translucent ISM, where ZG24 observe a decrease in $R(V)$ with increasing $A(V)$, the evolution of dust is dominated by accretion \citep{Hirashita12}, whereby metals from the gas phase are accumulated on the surface of dust grains. As a result, the growth rate of grain mass ($m$) is proportional to the grain surface area:
\begin{equation}
  \frac{d m}{d t} \propto 4\pi \rho a^2,
\label{eqn:accretion}
\end{equation}
where $\rho$ is the density of dust grains and $a$ is the ``effective radius,'' defined as $a \equiv (3m/4\pi \rho)^{1/3}$, a measurement of grain size. Factors affecting the accretion rate include the dust temperature, the metal abundance in the gas phase and the sticking probability \citep{Hirashita12}, all of which are nearly independent of $a$. If we further assume the mass density of a dust grain, $\rho$, to be independent of its effective radius $a$, then $\frac{d a}{d t}$ is independent of effective radius $a$. Therefore, the net growth of effective radius, $\Delta a$, over some timespan is roughly the same for all dust grains despite of their different initial sizes. 

\begin{figure}
    \centering
    \includegraphics[width=1.0\linewidth]{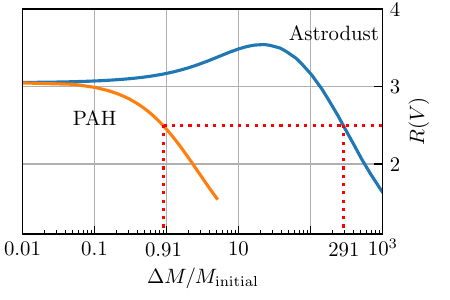}
    \caption{$R(V)$ variation as a function of the mass accreted onto astrodust or PAHs relative to initial mass, holding the other component (PAHs or astrodust) fixed. For $R(V)$ to decrease from $3.1$ to $2.5$, the astrodust mass must grow by a factor of $\sim 291$. This greatly exceeds the available mass budget of Mg, Fe, and Si (the major components of astrodust) in the gas. However, PAHs only have to grow by $91\%$ in mass, or 54~ppm of C, which is a viable amount given the mass budget of carbon in the gas phase \citep{Jenkins09, HD21_depletion}.  }
    \label{fig:astrodust_accretion}
\end{figure}

Optical-NIR extinction of the astrodust component is dominated by the grains with effective radius of $\gtrsim 0.1\,\mathrm{\mu m}$ \citep{Astrodust_dielectric, Astrodust_PAH}. As illustrated in Figure~\ref{fig:astrodust_accretion}, in order for accretion onto astrodust to effectively change $R(V)$ from 3.1 to 2.5, $\Delta a$ has to reach $\simeq 0.3\,\mathrm{\mu m}$. This would require $\simeq 291$ times the initial mass of astrodust to be accreted onto the astrodust grains, greatly exceeding what is available in the gas phase. The abundance of major elements of astrodust, Mg, Fe and Si, available in gas phase is 7.1, 0.88 and 6.6 parts per million hydrogen atoms (ppm) \citep{Jenkins09, HD21_depletion}, respectively, which only enables growth of dust mass by $\sim 16\%$, $2\%$ and $17\%$. This rules out the possibility that accretion onto large grains could change $R(V)$ by the observed amount.

On the other hand, the red wing of the 2175\,\AA{} feature extends to optical-NIR wavelengths and has a profile that is steep enough to change the slope of the extinction curve in this regime \citep[Figure~3]{Astrodust_PAH}. Arising from $\pi^* \leftarrow \pi$ electronic transitions in carbon rings, the strength of the feature is proportional to the total mass of PAHs, since their physical size (nanometers) is much smaller than 2175\,\AA\ (i.e., absorption is in the electric dipole limit). We can estimate $R(V)$ variation as a function of PAH abundance, represented by the total dust mass fraction of PAHs ($\qpah$). In Figure~\ref{fig:astrodust_accretion}, the orange curve shows $R(V)$ variation with PAH growth, holding the astrodust component fixed. In order to change $R(V)$ by $\sim 0.6$, PAH mass only needs to grow by 91\%, corresponding of an increase in $q_\mathrm{PAH}$ from $6\%$ to $11\%$. This amount of growth requires another 54~ppm of C to be added to PAHs, while there is $\sim$200~ppm available in the gas phase \citep{Jenkins09, HD21_depletion}. 

Although allowed by the mass budget, the question then arises as to whether PAHs can double in mass within a reasonable timescale, which we take the lifespan of a typical molecular cloud as a lower limit.  Unfortunately, the growth pattern of PAHs in molecular clouds remains unclear. Inspired by \citet{Krasnokutski_PAH_accretion}, we estimate the time scale by a simple accretion model. We assume carbon atoms in the gas that collide with the rim (width $\approx 1.4\,\mathrm{\AA}$) of a PAH molecule are incorporated into the molecule. We assume atoms in gas follow a Maxwellian velocity distribution with $T=20K$, and calculate the time scale for a few particular PAH molecules to double in size, tabulated in Table~\ref{tab:timescale}. 

The resulting timescales of only $\sim$2\,Myr are much shorter than the lifespan of a molecular cloud (typically $10\sim20\,\mathrm{Myr}$). The accretion rate of carbon ions onto neutral PAHs could be enhanced by more than an order of magnitude due to electromagnetic effects \citep{Weingartner99_cross_section}.
We describe this estimate in detail in Appendix~\ref{app:PAH_accretion}. We therefore conclude that doubling the PAH mass is plausible.

Finally, we emphasize that our result is not specific to the astrodust model. Although the precise range of $R(V)$ variation may change with different dust models, the trend of decreasing $R(V)$ with PAH mass ratio holds, because it only relies on PAHs being the carrier of the 2175~\AA{} feature.

\newpage
\section{Discussion} \label{sec:discussion}

\subsection{Constraints on the PAH evolution}

Our findings underscore the importance of the growth of PAHs in the ISM, which is still poorly understood. To list a few questions:
\begin{itemize}
\vspace{-0.1cm}
\item Do PAHs form in-situ or migrate from denser regions? 
\vspace{-0.3cm}
\item How are carbon atoms ``assembled'' onto PAHs?
\vspace{-0.3cm}
\item What affects the growth or destruction rate of PAHs?
\vspace{-0.1cm}
\end{itemize}

PAH growth in the ISM has been suggested by various recent observations. \citet{Whitcomb24_PAH_Z_trend} found that with the decrease of metallicity, the power of the long-wavelength PAH emission features tends to decrease, while the power of the short-wavelength features increases. \citet{Whitcomb24_PAH_Z_trend} attribute this to the inhibited grain growth at low metallicity. Type Ia supernovae are observed to have anomalously low $R(V)$ \citep{Popovic23_dust_SNe}, which could be connected to PAHs if, e.g., PAHs are formed in supernova shocks \citep{Tielens08_PAH_review}.
Aromatic molecules can also be formed in a ``bottom-up'' way from precursors (e.g., \citealt{Jones11_PAH_formation, McGuire21_PAH_detection}).
 Experiments and computations indicate that PAHs can acquire carbon from the gas over a wide temperature range \citep{Krasnokutski_PAH_accretion}. 

Taken together with the results of the present study, these observations suggest that PAH growth in the ISM is a major component of the PAH life cycle in galaxies and is responsible for much of the observed PAH mass.

\subsection{Implication on future JWST observations}

JWST/MIRI covers PAH emission features at 6.2, 7.7, 8.6, 11.2, 12.7 and 16.4 $\mu m$ and is powerful in determining PAH composition (e.g. \citealt{Chastenet23_PHANGS, PAH_Seyfert, Sturm24_JWST_PAH}). JWST/MIRI MRS measurements of a series of targets in the transluscent ISM with similar $A(V)$ but a gradient in $R(V)$ could trace the different evolutionary phases of PAHs. In addition to testing the hypothesis put forward here, this could provide valuable constraints on the carbon cycle in the ISM.

$R(V)$ variation caused by the 2175~\AA{} feature relies primarily on the total mass ratio of PAHs, and is not sensitive to the PAH size distribution. Nevertheless, $R(V)$ maps indicate locations of significant variation in PAH abundance, which are potential sites of growth, destruction, or migration of PAHs. Therefore, these maps can be used to identify optimal targets for observing PAH evolution with JWST.

\section{Summary}

In this work, we find an anti-correlation between $R(V)$ and $\qpah$ in the translucent ISM. We observe this phenomenon both in the vicinity of various molecular clouds in the Milky Way and in the LMC. These regions are also where $R(V)$ decreases with $A(V)$ in ZG24.

We propose a theory to explain both anti-correlations: $R(V)$ variation is driven by the growth of PAHs. The 2175\,\AA{} feature enables PAHs to change the slope of extinction curves at optical-NIR wavelengths even if the sizes of the molecules are far smaller than the wavelength of the B or V band. This theory is consistent both with the mass budget of carbon atoms available in gas phase and with the estimated time scale for PAHs to gain the needed mass. 

We discuss potential JWST observations of the translucent ISM to resolve PAH absorption and emission features. Such observations would not only test the PAH-$R(V)$ correlation but also impose important constraints on the evolution of PAHs and the carbon cycle in the ISM.

\section*{Acknowledgments}
We would like to thank Serge Krasnokutski and Thomas Henning for discussions on PAH growth. XZ and GG were supported by a Humboldt Foundation Sofja Kovalevskaja Award granted to GG. This research was carried out in part at the Jet Propulsion Laboratory, California Institute of Technology, under a contract with the National Aeronautics and Space Administration (80NM0018D0004).

\appendix

\section{2D $R(V)$ maps of clouds in the Milky Way}
\label{app:binning}

In order to compare the $R(V)$ results with the 2D $q_\mathrm{PAH}$ map, we build $A(V)$ and $R(V)$ maps using binning method. We select stars that pass the \texttt{basic cut} and the \texttt{$T_\mathrm{eff}$ confidence} cut in \cite{ZG24a}. Because the clouds we select are the main source of extinction along the lines of sight, we can simply remove the foreground contamination by only using stars beyond the minimal distances of dust clouds, as tabulated in Table~\ref{tab:k_b_clouds}, determined by \cite{Schlafly14_cloud_distance}.

The \textit{Gaia} XP catalog has a magnitude limit of $17.65$ mag \citep{GaiaXPProcessingValidation,GaiaXPExternalValidation}, which excludes heavily-extincted stars behind some dust clouds so that binning method could only measure partial extinction. To make sure our $R(V)$ map describe the majority of dust along the line of sight, we select stars by comparing extinction measure by \textit{Gaia} XP, $A(V)_\mathrm{XP}$, with that of 2D extinction map ``SFD'' \citep{SFD}, $E_\mathrm{SFD}$:

\begin{equation}\label{eqn:sfd_cut}
A(V)_\mathrm{XP} - k E_\mathrm{SFD} - \texttt{bias} >-0.5\,\mathrm{mag}.
\end{equation}
Coefficients $k$ and $\texttt{bias}$ are fitted using low-extinction regions of the cloud with $0.1<E_\mathrm{SFD}<0.8\, \mathrm{mag}$, where most of the XP stars are behind clouds.  We tabulate the coefficient values in Table~\ref{tab:k_b_clouds} .

We allocate stars into bins of $\Delta \ell\times \Delta b=0.1^\circ\times0.1^\circ$, and calculate the average $A(B)$, $A(V)$ and $R(V)$ by the following equation in each bin:

\begin{equation}\label{eqn:weight_of_stars}
\begin{split}
&\bar{A}(B) = \sum_i w_{B, i} A(B)_i /\sum_i w_{B, i}, \\
&\bar{A}(V) = \sum_i w_{V, i} A(V)_i /\sum_i w_{V, i}, \\
&\bar{R}(V) = \bar{A}(V)/(\bar{A}(B)-\bar{A}(V) )
\end{split}
\end{equation}
where $i$ is the index of stars in each bin, and the weight of each star is given by $w_i = (\sigma(A(B))^2+\sigma(A(V))^2+0.01^2)^{-1}$.

\begin{table}
    \centering
    \begin{tabular}{cccccc}
        Cloud name & \texttt{Aquila S} & \texttt{California} & \texttt{Orion A} & \texttt{Perseus} & \texttt{Ursa Ma}\\
        \hline
        Minimal distance (kpc) & 0.05 & 0.3 & 0.3 & 0.2 & 0.2\\
        \hline
        $k$& 2.14 & 1.78 & 1.92 & 1.87 & 2.19\\
        \hline
        $\texttt{bias}$ & 0.17 & 0.24 & 0.12 & 0.21 & 0.14 \\
        
    \end{tabular}
    \caption{Coefficients to select stars into the 2D $R(V)$ maps.}
    \label{tab:k_b_clouds}
\end{table}

\section{2D $R(V)$ maps for LMC and SMC}
\label{app:MC_binning}

We select the stars using the same standard as Appendix~\ref{app:binning}. In order to compare with the $q_{PAH}$ maps from \cite{Chastenet19}, which remove foreground emission, we also remove the foreground extinction of LMC and SMC.  We still select stars by the quality cuts described in Appendix~\ref{app:binning}, but additionally define two group of stars: ``foreground'' and ``target'' stars. The foreground stars are the stars in the Milky Way halo that fulfill \textbf{both} of the following condition:
\begin{equation}
\begin{split}
&\varpi_\mathrm{est} - 2\sigma(\varpi_\mathrm{est})>1/(5\mathrm{kpc}) \\
&|\sin(b)| \left(\varpi_\mathrm{est} + 2\sigma(\varpi_\mathrm{est})\right)<1/(0.4\mathrm{kpc}) 
\end{split}
\end{equation}
We use binning method (Equation~\ref{eqn:weight_of_stars}) calculate extinction contributed by the dust in the Milky Way disk and halo. 

We select ``target'' stars in the LMC or SMC by choosing stars that fulfill \textbf{either}:
\begin{equation}
\varpi_\mathrm{est} + 2\sigma(\varpi_\mathrm{est})<1/(30\mathrm{kpc}),
\end{equation}
\textbf{or}
\begin{equation}
\begin{split}
&\varpi_\mathrm{est} - \sigma(\varpi_\mathrm{est}) < 1/(30\mathrm{kpc}); \\
&(\mu_{\alpha} - \bar{\mu}_{\alpha})^2 + (\mu_{\delta} - \bar{\mu}_{\delta})^2 < (0.8 \mathrm{mas/yr})^2.
\end{split}
\end{equation}
$\bar{\mu}_{\alpha}$ and $\bar{\mu}_{\delta}$ are the central proper motion. We use $\bar{\mu}_{\alpha}=1.91\,\mathrm{mas/yr}$ and $\bar{\mu}_{\delta}=0.229\,\mathrm{mas/yr}$ for LMC and $\bar{\mu}_{\alpha}=0.772\,\mathrm{mas/yr}$ and $\bar{\mu}_{\delta}=-1.117\,\mathrm{mas/yr}$ for SMC \citep{Kallivayalil13_MCpm}. Proper motion of stars are from \textit{Gaia} DR3 \texttt{pmra} and \texttt{pmdec}.
We calculate $R(V)$ map of LMC or SMC by the ``target'' stars using Equation~\ref{eqn:weight_of_stars}, but using the residual $A(V)$ and $A(B)$ subtracting the foreground values.

\section{Estimation of time scale for PAH accretion}
\label{app:PAH_accretion}
In this appendix, we estimate the time scale of PAH growth by accretion. Based on \citet{Krasnokutski_PAH_accretion}, we assume that C atoms that collide with the rim of PAH molecules are incorporated into the molecule. We assume the width of the rim is the spacing between C atoms in PAHs, $\Delta R \approx 1.4\,\mathrm{\AA}$. The number of atoms collected per unit time by a PAH molecule (assumed to be circular) is:
\begin{equation}\label{eqn:dndt}
  \frac{\mathrm{d} N}{\mathrm{~d} t}
  = \frac{1}{2} \int_{-1}^{+1} \hspace{-1.3em} \mathrm{~d}(\cos \theta)
  \int_0^{\infty} \hspace{-0.9em} \mathrm{d} v~p(v)~2\pi R \Delta R|\cos\theta| n v
  = \pi R \, \Delta R \, n \langle v\rangle,
\end{equation}
where $N$ is the number of carbon atoms in the PAH molecule, $\theta$ is the inclination angle, $R$ is the radius of the PAH molecule, $n$ is the number density of carbon atoms in gas, $v$ is the speed of the carbon atoms, and $p(v)$ is the probability of speed $v$. If the 3D velocity follows Maxwellian distribution, then we have
\begin{equation}
  \langle v\rangle=\sqrt{\frac{8 k_B T}{\pi m}},
\end{equation}
where $T$ is the temperature of the carbon atoms, and $m=12\,\mathrm{Da}$ is the atomic mass of C. 

The radius of a compact pericondensed PAH molecule is related to number of carbon atoms by:
\begin{equation}
  \pi R^2 \approx N \left(\Delta R\right)^2.
\end{equation}

We can use the above to formulate Equation~\ref{eqn:dndt} in terms of $N$ (instead of $R$), resulting in
\begin{equation}
\frac{\mathrm{d} N}{\mathrm{~d} t} = (\Delta R)^2 n \sqrt{\frac{8 k_B T N}{m}} \, .
\end{equation}
The solution to this differential equation is given by
\begin{equation}
N = \left((\Delta R)^2 n t\  \sqrt{\frac{ 2 k_B T }{m}} + \sqrt{N(t=0)}\right)^2
\end{equation}

If $T=20\,\mathrm{K}$ and $n=10^{-2}\, \mathrm{cm^3}$, we find time scale for $\sqrt{N}$ to increase by 1 is:
\begin{equation}
 \left((\Delta R)^2 n \  \sqrt{\frac{ 2 k_B T }{m}}\right) = 1.0\,\mathrm{Myr^{-1}}
\end{equation}


\begin{table}
    \centering
    \begin{tabular}{c|c}
       Typical types & Time scale of accretion  \\
        of PAH  & to double the mass (Myr) \\
       \hline
        Naphthalene ($\mathrm{C_{10}H_{8}}$) & 1.28 \\
        \hline
        Phenalene ($\mathrm{C_{13}H_{10}}$) & 1.46 \\
        \hline
        Pyrene ($\mathrm{C_{16}H_{10}}$) & 1.61 \\
        \hline
        Corannulene ($\mathrm{C_{20}H_{10}}$) & 1.80 \\
        \hline
        Coronene ($\mathrm{C_{24}H_{12}}$) & 1.98 \\
    \end{tabular}
    \caption{Time scales for typical types of PAH to double their mass by accretion. These time scales are much shorter than the lifespan of molecular clouds ($\simeq20\ \mathrm{Myr})$, such that growth by accretion is a viable means of explaining the variation of PAH abundance that give rise to the $R(V)$ variation observed in ZG24.}
    \label{tab:timescale}
\end{table}

Various factors could affect the time scale. To name a few, sticking probability is assumed to be $100\%$ in our estimate, but in reality should depend on factors such as binding energy, temperature, inclination, velocity and angle. Accretion efficiency also decreases with the depletion of carbon atoms in gas. Because a fraction of PAHs is ionized, Coulomb focusing could accelerate accretion \citep{Weingartner99_accretion}. 

\bibliography{reference}

\end{document}